\documentclass[twoside,reqno]{heron}
\usepackage{epsfig,cite,colordvi}
\usepackage{graphicx}
\usepackage{url}
\usepackage{amssymb,amsmath,amscd,epsf}
\usepackage{times}
\usepackage{makeidx}
\begin{document}

\title{Why do Nilsson quantum numbers remain good at moderate deformations?}

\runningheads{Nilsson quantum numbers at moderate deformations}{D. Bonatsos, I.E. Assimakis, 
A. Martinou, S. Peroulis, S. Sarantopoulou}

\begin{start}

\author{Dennis Bonatsos}{1}, \coauthor{I.E. Assimakis}{1}, \coauthor{Andriana Martinou}{1}, 
\coauthor{S. Peroulis}{1}, \coauthor{S. Sarantopoulou}{1}, \coauthor{N. Minkov}{2}

\index{Bonatsos, D.}
\index{Assimakis, I.E.}
\index{Martinou, A.}
\index{Peroulis, S.}
\index{Sarantopoulou, S.}
\index{Minkov, N.}

\address{Institute of Nuclear and Particle Physics, National Centre for Scientific Research 
``Demokritos'', GR-15310 Aghia Paraskevi, Attiki, Greece}{1}

\address{Institute of Nuclear Research and Nuclear Energy, Bulgarian Academy of Sciences, 72 Tzarigrad Road, 1784 Sofia, Bulgaria}{2}

\begin{Abstract}

The Nilsson model is a simple microscopic model which has been extensively used over the years for the interpretation of a bulk of experimental results. The single particle orbitals in this model are labeled by quantum numbers which are good in the limit of large nuclear deformations. However, it is generally admitted that these quantum numbers remain good even at moderate deformations. We show that this fact is due to the existence of an underlying approximate symmetry, called the proxy-SU(3) symmetry. The implications of proxy-SU(3) on various aspects of nuclear structure will be discussed.

\end{Abstract}
\end{start}


\section{Introduction}

The Nilsson model \cite{Nilsson1,Nilsson2} is an elementary shell model, taking into account the spin-orbit interaction, which is instrumental in creating the nuclear magic numbers \cite{Ring}. In addition, it takes into account the tendency of nuclei to acquire axially symmetric deformed shapes \cite{BM}, since it is based on a three-dimensional (3D) harmonic oscillator with cylindrical symmetry, i.e., with two equal frequencies in the axes $x$ and $y$ and a different one in the $z$ axis. Despite its simplicity, the Nilsson model has been extremely useful 
over several decades in the interpretation of a bulk of experimental findings in the study of atomic nuclei
\cite{Casten}. 

The notation \cite{Nilsson2} adapted since its early days for the energy levels of the Nilsson model makes use of the quantum numbers $K[N n_z \Lambda]$, where $N$ is the total number of oscillator quanta, $n_z$ is the number of quanta along the cylindrical symmetry axis $z$, $\Lambda$ is the projection of the orbital angular momentum on the $z$ axis, and $K$ is the projection of the total angular momentum along the $z$ axis. The last two quantities are connected to the projection of the spin on the $z$ axis, $\Sigma$, by 
$K=\Lambda+\Sigma$. These quantum numbers are known to be good in the asymptotic limit of large deformations, in which the spin-orbit term in the Hamiltonian can be ignored \cite{Nilsson2}. However, they have been used over the years for labeling the Nilsson energy levels also at moderate deformations, and even at small ones \cite{Casten}. 

The fact that the Nilsson quantum numbers remain a reasonable tool for labeling the energy levels far away from the asymptotic region in which they are mathematically justified, implies that there must be a physical and/or mathematical reason behind their success. In the present article we point out that an approximate symmetry, called the proxy-SU(3) symmetry \cite{proxy1,proxy2}, can explain this success. The main features of the proxy-SU(3) symmetry will be discussed and some of its consequences in the behavior of even-even atomic nuclei will be reviewed. 

\section{The proxy-SU(3) symmetry scheme}

The proxy-SU(3) symmetry is based on the fact that Nilsson orbitals characterized by quantum numbers differing by $\Delta K[\Delta N \Delta n_z \Delta\Lambda]=0[110]$ possess very large spatial overlaps
\cite{Karampagia}, thus favoring large interaction between them. The large size of the overlaps is due to the fact that the two orbitals posses the same values of the angular momentum and spin quantum numbers 
$\Lambda$, $K$, $\Sigma$, 
while they only differ by one unit in the number of quanta along the $z$ axis. Loosely speaking, they represent two ellipsoids having the same orientation in space, with one of them being slightly more elongated than the other. 

The importance of 0[110] pairs has been first realized in relation to the proton-neutron interaction,
by studying double differences of nuclear masses \cite{Cakirli1,Cakirli2}. It was then realized that the same mathematical reasons leading to maximal proton-neutron interaction between 0[110] pairs of orbitals, can lead to maximal interaction in the cases of proton-proton pairs and neutron-neutron pairs as well
\cite{proxy1}. 

In particular, the great similarity between 0[110] pairs can lead to an approximate restoration of the SU(3) symmetry of the 3D harmonic oscillator \cite{Wybourne}, which is valid up to the sd shell (20 protons or 20 neutrons) \cite{Elliott1,Elliott2}, but is broken in higher shells because of the spin-orbit interaction,
leading to the nuclear magic numbers 28, 50, 82, 126, 184, \dots instead of the 3D harmonic oscillator magic numbers 40, 70, 112, 168, 240, \dots \cite{Nilsson2}. 
Within each harmonic oscillator shell beyond 20 nucleons, a bunch of orbitals (the ones having the highest 
angular momentum) is pushed down in energy by the spin-orbit interaction, which acquires maximum values for these specific orbitals. But the shell losing these orbitals is invaded by another bunch of orbitals, coming down from the shell above for the same reason. 

The restoration of SU(3) is achieved by replacing the bunch of orbitals invading  a harmonic oscillator shell from above by the bunch of orbitals which deserted this shell going to the one below it \cite{proxy1}. In particular, each invading orbital is replaced by its 0[110] counterpart among the deserting orbitals.
Since the orbitals forming a 0[110] pair posses the same projections of total angular momentum, orbital angular momentum, and spin, the changes inflicted in the interaction matrix elements are few and small, 
as demonstrated in detail for all nuclear shells in Ref. \cite{proxy1}. Only one of the invading orbitals
(the one with the highest angular momentum) does not have a 0[110] counterpart and is therefore omitted. 
As a result, the ``new'' proxy-SU(3) shell can accommodate two nucleons less than the original nuclear shell \cite{proxy1}. 

It should also be pointed out that the invading orbitals possess parity opposite to the one of the deserting orbitals which they replace, since the two bunches of orbitals in discussion (invading and deserting) originate from adjacent harmonic oscillator shells with $\Delta N=1$. The consequences of this difference should be seriously taken into account if odd nuclei are to be considered. But for the study of even-even nuclei this difference has minimal effect, since only pairs of nucleons are present (and broken pairs are avoided),  which possess the same total parity irrespectively of the harmonic oscillator shell which they come from.  

From the above it is clear that the validity of the Nilsson labels $K[N n_z \Lambda]$ far away from the asymptotic limit of large deformations can be attributed to the existence of the approximate proxy-SU(3) symmetry \cite{proxy1} underlying the nuclear shells above 28 nucleons. In other words, the SU(3) symmetry 
of the harmonic oscillator in the absence of spin-orbit interaction, which is valid in the limit of large deformations, is proven to survive at moderate deformations in the form of the approximate proxy-SU(3) symmetry, therefore allowing the quantum numbers which are good at large deformations to remain useful 
also at moderate deformations.   

The validity of the Nilsson labels is not the sole consequence of the existence of the proxy-SU(3) symmetry. A few effects due to the proxy-SU(3) symmetry will be reviewed below.

\section{Particle-hole symmetry breaking} 

An immediate consequence of the existence of the proxy-SU(3) symmetry is the breaking of particle-hole symmetry within a nuclear shell \cite{proxy2,37J,40GC}. We will show how this occurs.
  
The $n$-th shell of the 3D harmonic oscillator is known to posses a unitary symmetry U($(n+1)(n+2)/2$)
\cite{Wybourne}, possessing  an SU(3) subalgebra \cite{BK}. 
For a given number of neutrons (or protons) within a nuclear shell, many states can occur. These are classified using the irreducible representations (irreps) of the underlying SU(3) subalgebra, 
labelled, using the Elliott notation \cite{Elliott1,Elliott2}, by $(\lambda,\mu)$. The irreps of SU(3) occurring within U(N) for a given number of particles can be determined by standard codes \cite{code}.
Among all these irreps, the configuration (the state) which is most probable to appear is the one corresponding in group theoretical language to the highest weight irreducible representation (h.w. irrep) of SU(3). The codes just mentioned list the irreps in order of decreasing weight, thus the problem of 
determining the appropriate irrep for any number of nucleons in any shell is also solved. Lists of the h.w. irreps occurring in nuclear shells can be found in Refs. \cite{proxy2,37J,40GC}. A partial list is given here in Table 1. 

From Table 1 one can see that particle-hole symmetry is broken. Let us briefly discuss the origins of this effect. In fact, everything can be derived using only the Pauli principle and the short range nature of the nucleon-nucleon interaction \cite{Ring,Casten}.
The short range nature of the nucleon-nucleon interaction \cite{Ring,Casten} favors maximal spatial overlaps, which occur in the case of symmetrized spatial wave functions.
Because of the Pauli principle, the spin-isospin part of the wave function has to be antisymmetric, in order to guarantee the antisymmetric character of the total wave function. 
Looking into the details, one can see that the highest weight irrep is the irrep with the maximum spatial symmetrization possible, given the restrictions imposed by the Pauli principle. One needs the 
Gelfand-Tsetlin diagrams to show this \cite{MartinouPhD}.
One has to realize that the influence of the Pauli principle is not exhausted by imposing the antisymmetry of the spin-isospin part of the wave function. It does influence in parallel the structure of the spatial part of the wave function as well. 

The main conclusion so far is that the particle-hole symmetry breaking \cite{proxy2,37J,40GC} comes from the Pauli principle and the short range of the nucleon-nucleon interaction alone. From the mathematical point of view, the process is ``mechanical'', but it is based on two basic physical principles. 
There is no need for a specific Hamiltonian in order to do the above, since any nuclear Hamiltonian has to respect these two basic principles.  

The fact that the quadrupole-quadrupole interaction is maximized in deformed nuclei \cite{Casten} is due to the two basic physical principles mentioned above. Actually this happens only in the first half of the shell. In the second half of the shell it is NOT the irrep giving the highest
quadrupole-quadrupole interaction (equivalently: the highest eigenvalue of the second order Casimir operator of SU(3)) that is preferred by the Pauli principle and the short range interactions
\cite{proxy2,37J,40GC}. 
Although this has been known since many years, as it can clearly be seen in Table 5 of 
Ref. \cite{pseudo1}, it seems that it has not received adequate attention so far.  One of the reasons might be that the particle-hole asymmetry does not appear in the sd shell studied by Elliott, 
except for 7 particles (\cite{Elliott1}, see also Table 1 of the present paper), thus it went unnoticed. 

The ideas discussed above can be clarified through an example.
Consider the U(15) shell (proxy-SU(3) sdg shell). If one has 10 particles in it, the h.w. irrep is (20,4), 
as seen in Table 1. If one looks at 10 holes, i.e. 30-10=20 particles, the h.w. irrep is (20,0), while in the case of particle-hole symmetry one would have expected to obtain the conjugate irrep of (20,4), i.e. (4,20). One can just plot the (20,0) and (4,20) irreps in order to see which looks more symmetric. The Young diagram for (20,0)  consists of 20 boxes in the first row and no boxes in the second row. The Young diagram for (4,20) has 24 boxes in the first row and 20 boxes in the second row. We know that by definition  boxes in the same row mean symmetrization, while boxes in the same column mean antisymmetrization. Therefore (20,0) is purely symmetric, while (4,20) contains lots of antisymmetrizations.  The most symmetric irrep gives the highest spatial overlaps, and it corresponds to the most antisymmetric spin-isospin irrep. The (4,20) irrep does have a higher Casimir eigenvalue than the (20,0) irrep, but it is not the one preferred by the short range interaction in combination with the Pauli principle. And all this happens because the nucleon-nucleon interaction is of short range, thus it prefers the most symmetric spatial irreps. 

In the  case of the sd shell of protons only, or neutrons only, the shell can accommodate 12 particles.  Then from Table 1 for 5 particles one gets the (5,1) irrep, but for 7 particles one gets the (4,2) irrep, not the (1,5) irrep. This result already exists in the original Elliott papers \cite{Elliott1}.

In conclusion, the physical principles leading to the final results are the same both in the first half and in the second half of a given shell. In both cases the interaction is short range, thus the spatial irrep still  has to be most symmetric. The Pauli principle is there in both cases, thus the spin-isospin irrep still has to be most antisymmetric. 
In both cases the highest weight SU(3) irrep is preferred. The only difference is that while in the first half of the shell the h.w. irrep coincides with the irrep possessing the highest eigenvalue of the second order Casimir operator of SU(3), this is not any more the case in the second half of the shell, resulting 
in the particle-hole symmetry breaking.

\section{Prolate to oblate shape/phase transition} 

An immediate consequence of the particle-hole symmetry breaking is the occurrence of a shape/phase transition 
from prolate to oblate shapes just below the shell closures of $Z=82$ and $N=126$. This effect has been adequately discussed in Refs. \cite{proxy2,proxy3,37J}, to which the reader is addressed. 

A question which could be raised is if and why the proxy-SU(3) scheme could be used near the shell borders.
In this direction one could remark that when one considers ground state properties only, the Elliott SU(3) is just a classification scheme, which works for all nuclei, deformed or not deformed, as seen in the original Refs. \cite{Elliott1,Elliott2,Elliott3}. The O(6) basis could also be used instead, if the relevant group theoretical work had been done, and the results would have been equally good. Notice that the O(6) classification of the sd shell already exists in the first Elliott paper, as an appendix \cite{Elliott1}. If one tries to describe spectra, then the differences show up. In SU(3) on would get good spectra only for deformed nuclei.  

\section{Shape coexistence}\label{coex}

Shape coexistence \cite{odd,Wood,Heyde} in even nuclei occurs when a ground state band  
based on the $0_1^+$ ground state is accompanied by a low lying $0^+$ band of significantly different deformation and clearly different structure. Good examples are provided by the $_{82}$Pb and $_{80}$Hg isotopes, as well as by the $_{50}$Sn isotopes \cite{odd,Wood,Heyde}. The existence of the band with different deformation is attributed to two-particle--two-hole (2p-2h) excitations across the proton shell gaps 82 and 50 respectively. 
A detailed map of regions of coexistence on the nuclear chart has been provided by a recent review (see Fig. 8 of Ref. \cite{Heyde}). However, the limitation of shape coexistence within specific regions of the nuclear chart remains an open problem. 

On the other hand, trying to understand the new features of nuclear magic numbers far from stability, 
which become gradually available through the advent of radioactive ion beam facilities and the steadily improved sensitivity of the experiments, one has to take into account the competition between spin-orbit--like magic numbers (14, 28, 50, 82, 126, \dots) and 
harmonic oscillator magic numbers (2, 8, 20, 40, 70, 112, \dots) \cite{Sorlin}. 
The competition between  spin-orbit--like magic numbers and harmonic oscillator magic numbers can also be seen in a calculation in the framework of the Nilsson model \cite{Nilsson1,Nilsson2},
similar to the one performed in Ref. \cite{proxy1} for justifying the approximation implemented in the proxy-SU(3) scheme. In particular one can see that while the usual nuclear magic numbers prevail at small deformations, the harmonic oscillator magic numbers become dominant at large deformations, while at intermediate deformations the gaps appearing at both sets of magic numbers are of comparable size. 
The results of this calculation will be presented elsewhere \cite{Assimakis}.  

As a consequence, for each nucleus there are two competing irreps, the one coming from the proxy-SU(3) scheme, which can be determined from the tables given in Refs. \cite{proxy2,37J}, and the one coming from the 3D harmonic oscillator, for which the relevant irreps can also be found from the tables given in Refs. \cite{proxy2,37J},
as we will explain below through two examples. Table 1 of the present paper contains all necessary results.

a) Consider 100 nucleons. Within the proxy-SU(3) pfh shell, which starts at 82 nucleons and possesses the 
U(21) symmetry, these correspond to 18 valence particles, thus the relevant h.w. irrep is (36,6), in agreement with Table 1. If we consider the 3D harmonic oscillator magic numbers, the 100 nucleons correspond to 30 valence particles within the 70-112 shell, which possesses the U(21) symmetry,
thus from Table 1 we see that the relevant h.w. irrep is (30,0). In both cases 
the number of h.o. quanta is $N_q=5$, thus the Casimir eigenvalues of the two irreps are directly comparable. 

b) Consider 120 nucleons. Within the proxy-SU(3) pfh shell, which starts at 82 nucleons and possesses the 
U(21) symmetry, having $N_q=5$, these correspond to 38 valence particles, thus the relevant h.w. irrep is (2,16), in agreement with Table 1. If we consider the 3D harmonic oscillator magic numbers, the 120 nucleons correspond to 8 valence particles within the 112-168 shell, which possesses the U(28) symmetry, having $N_q=6$, 
thus from Table 1 we see that the relevant h.w. irrep is (34,4).
In this case the number of quanta is different in the two shells involved, thus one has to take this into account when trying to compare the Casimir eigenvalues of the two irreps. 

In Refs. \cite{Castanos,Park} it has been found that the square of the deformation parameter $\beta$ is proportional 
to the square root of the second order Casimir operator of SU(3)\cite{IA}, 
 \begin{equation}\label{C2} 
 C_2(\lambda,\mu)= (\lambda^2+\lambda \mu + \mu^2+ 3\lambda +3 \mu). 
\end{equation}
Furthermore, the relevant equation for $\gamma$ reads \cite{Castanos,Park}
\begin{equation}\label{g1}
\gamma = \arctan \left( {\sqrt{3} (\mu+1) \over 2\lambda+\mu+3}  \right). 
\end{equation}

In Fig. 1 the quantity $\sqrt{C_2(\lambda,\mu)+3}$, which is proportional \cite{proxy2} to $\beta$,  and the deformation parameter $\gamma$ are plotted vs. the nucleon number (proton or neutron) $M$, for successive nuclear shells,
using the proxy-SU(3) scheme. The same quantities, calculated for the 3D harmonic oscillator (3D-HO),
having the SU(3) symmetry and  possessing the magic numbers 2, 8, 20, 40, 70, 112, 168, \dots, are shown for comparison. 

The following observations can be made. 

1) In each panel the curves corresponding to proxy-SU(3) and to the 3D-HO have the same shape, being displaced relative to each other just because they have different starting points. In the proxy-SU(3) the valence particles start being counted from the relevant nuclear magic number, while in the 3D-HO case 
the valence particles are enumerated from the relevant 3D-HO magic number. For example, in the second row of panels, valence particles for proxy-SU(3) start being counted from 50, while in the 3D-HO they start being enumerated from 40.  

2) For all nuclear shells shown, there is a region in which the values of  $\sqrt{C_2(\lambda,\mu)+3}$
predicted by the 3D-HO become lower than the proxy-SU(3) predictions. These are the regions 34-42, 60-72, 
96-116, 146-172. 
One should remember, however, that the relevant proxy-SU(3) and 3D-HO irreps belong to the same shell 
only below the relevant magic numbers of the 3D-HO (40, 70, 112, 168). Therefore one should focus attention 
to the regions 34-40, 60-70, 96-112, 146-168. The upper bounds of these regions 
coincide with the magic numbers of the 3D-HO, in striking agreement with the data reported in Ref. \cite{Ramos} for the Pb, Hg, and Pt isotopes. 
 
3) Most of the area of the regions found in 1) is covered by regions in which the $\gamma$ parameter 
predicted by the 3D-HO obtains values $\gamma \geq 30^{\rm o}$, corresponding to shapes on the oblate side
\cite{proxy2},
while proxy-SU(3) predicts $\gamma$ values on the prolate side, $\gamma < 30^{\rm o}$. 
The $\gamma \geq 30^{\rm o}$ regions are 34-40, 64-70, 104-112, 158-168. 

4) Within the regions of 2), one obtains from the proxy-SU(3) a prolate shape of higher deformation, 
while from the 3D-HO one obtains an oblate shape with lower deformation. 

Limiting ourselves to the part of the nuclear chart with $Z > 28$ and $N>28$, we see that the shape coexistence regions reported in Fig. 8 of Ref. \cite{Heyde} all fall within the regions determined above.
A more detailed comparison to the experimental observations in several regions where shape coexistence 
\cite{odd,Wood,Heyde} has been observed is deferred to  a longer publication.  

\section*{Acknowledgements}

Helpful discussions with K. Blaum, R. B. Cakirli, and R. F. Casten are gratefully acknowledged. 
Work partly supported by the Bulgarian National Science Fund (BNSF) under Contract No. DFNI-E02/6.

\begin{table}[htb]

\caption{Highest weight SU(3) irreps for U(n), n=6, 10, 15, 21, 28, 36. Irreps breaking the particle-hole symmetry are indicated by boldface. }
\smallskip
\small\noindent\tabcolsep=9pt

\begin{tabular}{ r l r r r r r r } 
\hline
\hline
\\[-8pt]

   &            & 8-20 & 28-50 & 50-82     & 82-126     &126-184&184-258\\
   &            & sd   & pf    & sdg       &  pfh       & sdgi  & pfhj  \\
M  & irrep      & U(6) & U(10) & U(15)     & U(21)      & U(28) & U(36) \\
 0 &            &(0,0) &(0,0)  &(0,0)      &(0,0)       & (0,0) & (0,0) \\  
 1 & [1]        &(2,0) & (3,0) & (4,0)     & (5,0)      & (6,0) & (7,0) \\
 2 & [2]        &(4,0) & (6,0) & (8,0)     &(10,0)      &(12,0) & (14,0)\\
 3 & [21]       &(4,1) & (7,1) &(10,1)     &(13,1)      &(16,1) & (19,1) \\
 4 & [$2^2$]    &(4,2) & (8,2) &(12,2)     &(16,2)      &(20,2) & (24,2)\\
 5 & [$2^2$1]   &(5,1) &(10,1) &(15,1)     &(20,1)      &(25,1) &(30,1) \\
 6 & [$2^3$]    &(6,0) &(12,0) &(18,0)     &(24,0)      &(30,0) & (36,0)\\
 7 & [$2^3$1&{\bf(4,2)}&(11,2) &(18,2)     &(25,2)      &(32,2) &(39,2)\\
 8 & [$2^4$]    &(2,4) &(10,4) &(18,4)     &(26,4)      &(34,4) & (42,4)\\
 9 & [$2^4$1]   &(1,4) &(10,4) &(19,4)     &(28,4)      &(37,4) &(46,4)\\
10 & [$2^5$]    &(0,4) &(10,4) &(20,4)     &(30,4)      &(40,4) & (50,4)\\
11 & [$2^5$1]   &(0,2) &{\bf(11,2)}&(22,2) &(33,2)      &(44,2) &(55,2) \\
12 & [$2^6$]    &(0,0) &{\bf(12,0)}&(24,0) &(36,0)      &(48,0) & (60,0)\\
13 & [$2^6$1]   &      &{\bf(9,3)} &(22,3) &(35,3)      &(48,3) &(61,3)\\
14 & [$2^7$]    &      &{\bf(6,6)} &(20,6) &(34,6)      &(48,6) &(62,6) \\
15 & [$2^7$1]   &      &{\bf(4,7)} &(19,7) &(34,7)      &(49,7) &(64,7) \\
16 & [$2^8$]    &      & (2,8) &{\bf(18,8)}&(34,8)      &(50,8) &(66,8)  \\
17 & [$2^8$1]   &      & (1,7) &{\bf(18,7)}&(35,7)      &(52,7) &(69,7)\\
18 & [$2^9$]    &      & (0,6) &{\bf(18,6)}&(36,6)      &(54,6) &(72,6) \\
19 & [$2^9$1]   &      & (0,3) &{\bf(19,3)}&(38,3)      &(57,3) &(76,3) \\
20 & [$2^{10}$] &      & (0,0) &{\bf(20,0)}&(40,0)      &(60,0) &(80,0) \\
21 & [$2^{10}$1]&      &       &{\bf(16,4)}&(37,4)      &(58,4) &(79,4)\\
22 & [$2^{11}$] &      &       &{\bf(12,8)}&{\bf(34,8)} &(56,8) &(78,8) \\
23 & [$2^{11}$1]&      &       &{\bf(9,10)}&{\bf(32,10)}&(55,10)& (78,10) \\
24 & [$2^{12}$] &      &       &{\bf(6,12)}&{\bf(30,12)}&(54,12) &(78,12) \\
25 & [$2^{12}$1]&      &       &{\bf(4,12)}&{\bf(29,12)}&(54,12) &(79,12) \\
26 & [$2^{13}$] &      &       &(2,12)     &{\bf(28,12)}&(54,12) &(80,12) \\
27 & [$2^{13}$1]&      &       &(1,10)     &{\bf(28,10)}&(55,10) &(82,10)  \\
28 & [$2^{14}$] &      &       & (0,8)     &{\bf(28,8)} &(56,8) &(84,8) \\
29 & [$2^{14}$1]&      &       & (0,4)     &{\bf(29,4)} &{\bf(58,4)} & (87,4)\\
30 & [$2^{15}$] &      &       & (0,0)     &{\bf(30,0)} &{\bf(60,0)}&(90,0) \\
31 & [$2^{15}$1]&      &       &           &{\bf(25,5)} &{\bf(56,5)} &(87,5) \\
32 & [$2^{16}$] &      &       &           &{\bf(20,10)}&{\bf(52,10)} &(84,10)\\
33 & [$2^{16}$1]&      &       &           &{\bf(16,13)}&{\bf(49,13)} &(82,13) \\
34 & [$2^{17}$] &      &       &           &{\bf(12,16)}&{\bf(46,16)} &(80,16)\\
35 & [$2^{17}$1]&      &       &           &{\bf(9,17)} &\bf{(44,17)} &(79,17)\\ 
36 & [$2^{18}$] &      &       &           &{\bf(6,18)} &{\bf(42,18)} &(78,18) \\
37 & [$2^{18}$1]&      &       &           &{\bf(4,17)} &{\bf(41,17)} &\bf{(78,17)} \\
38 & [$2^{19}$] &      &       &           &(2,16)      &\bf{(40,16)} &\bf{(78,16)} \\
39 & [$2^{19}$1]&      &       &           &(1,13)      &\bf{(40,13)} & \bf{(79,13)} \\
40 & [$2^{20}$] &      &       &           &(0,10)      &\bf{(40,10)} &\bf{(80,10)} \\
41 & [$2^{20}$1]&      &       &           &(0,5)       &\bf{ (41,5)} &\bf{(82,5)} \\
42 & [$2^{21}$] &      &       &           &(0,0)       & \bf{(42,0)} &\bf{(84,0)} \\

\hline
\end{tabular}
\end{table} 


\begin{figure}[htb]

{\includegraphics[width=55mm]{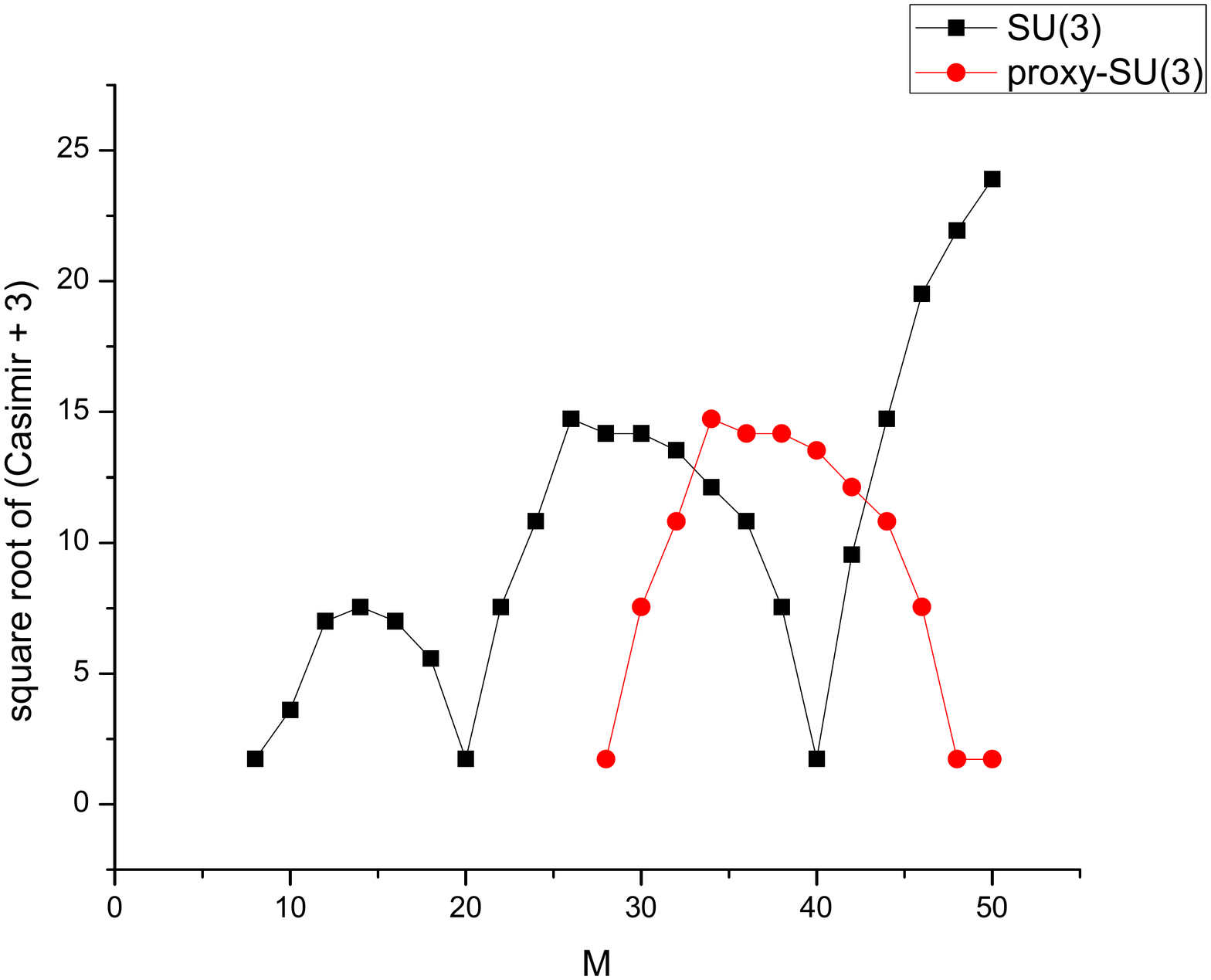}\hspace{1mm}
\includegraphics[width=55mm]{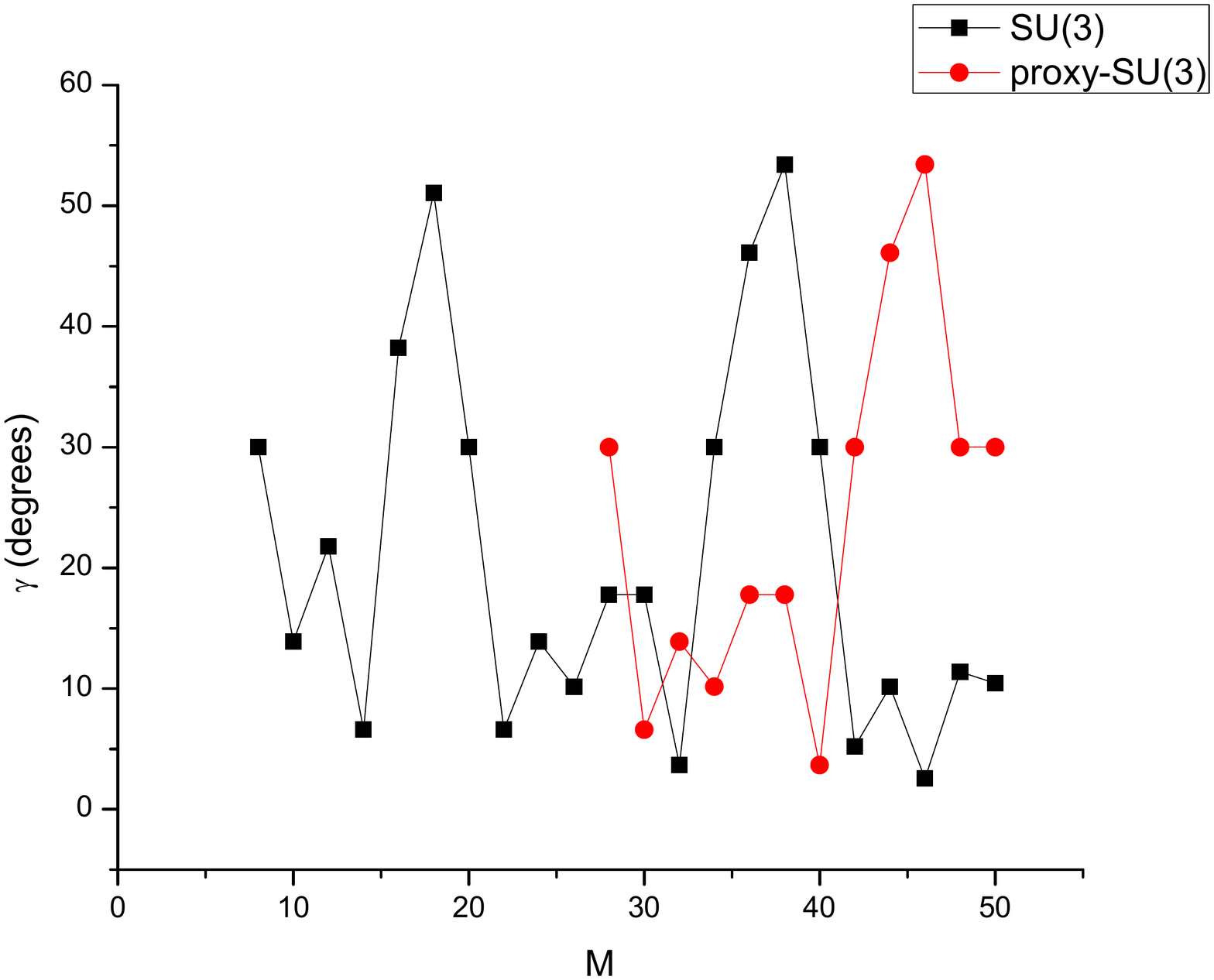}}
{\includegraphics[width=55mm]{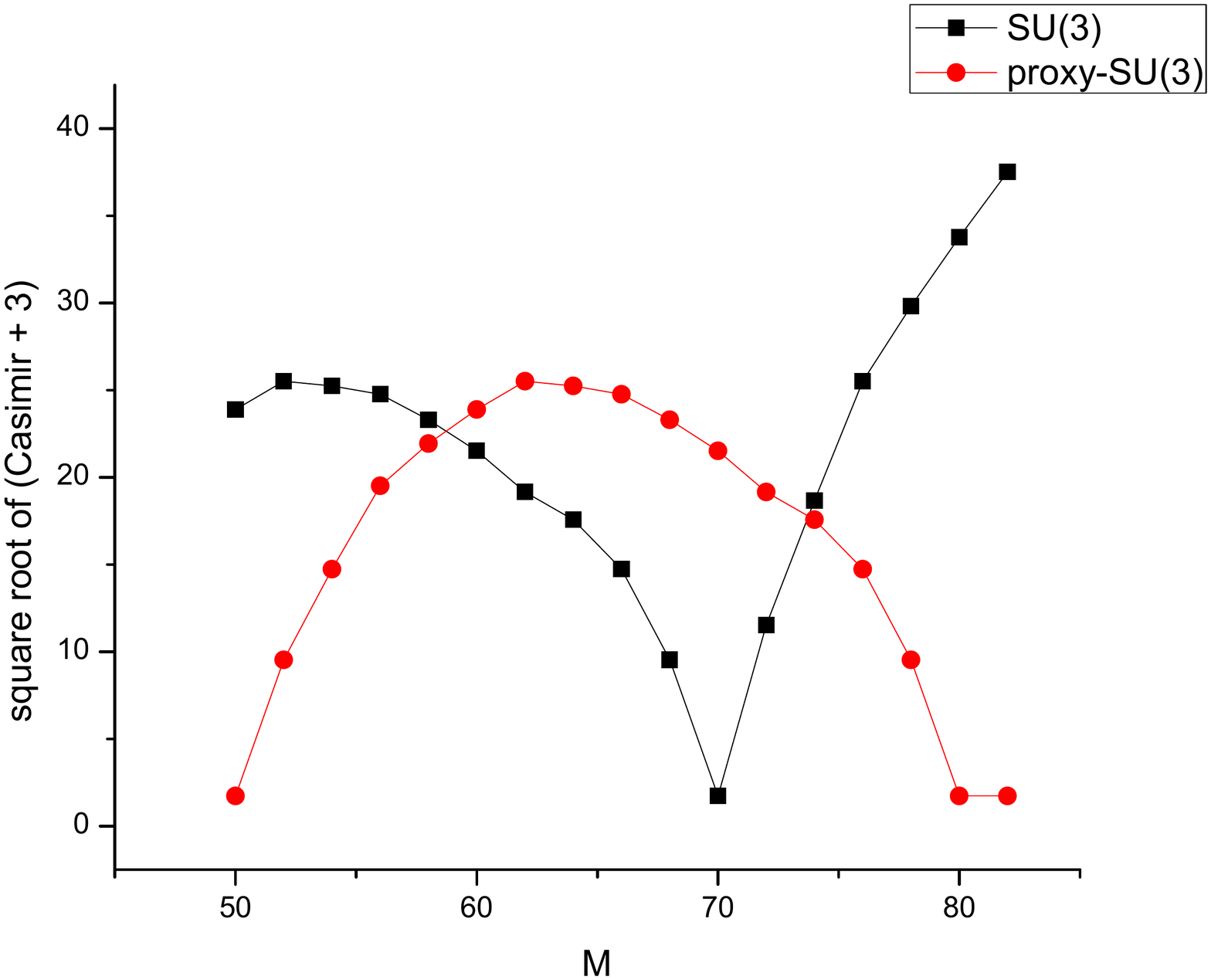}\hspace{1mm}
\includegraphics[width=55mm]{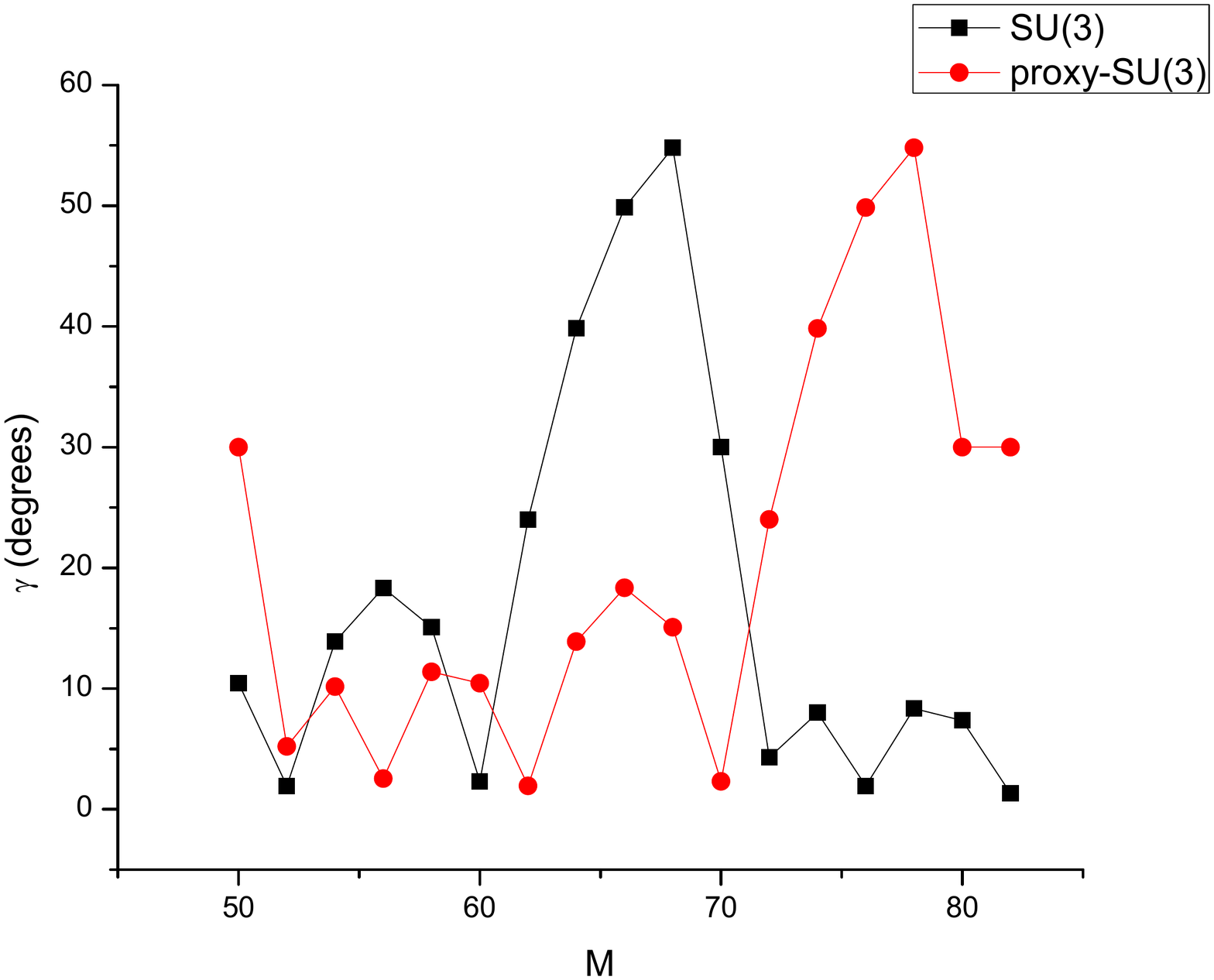}}
{\includegraphics[width=55mm]{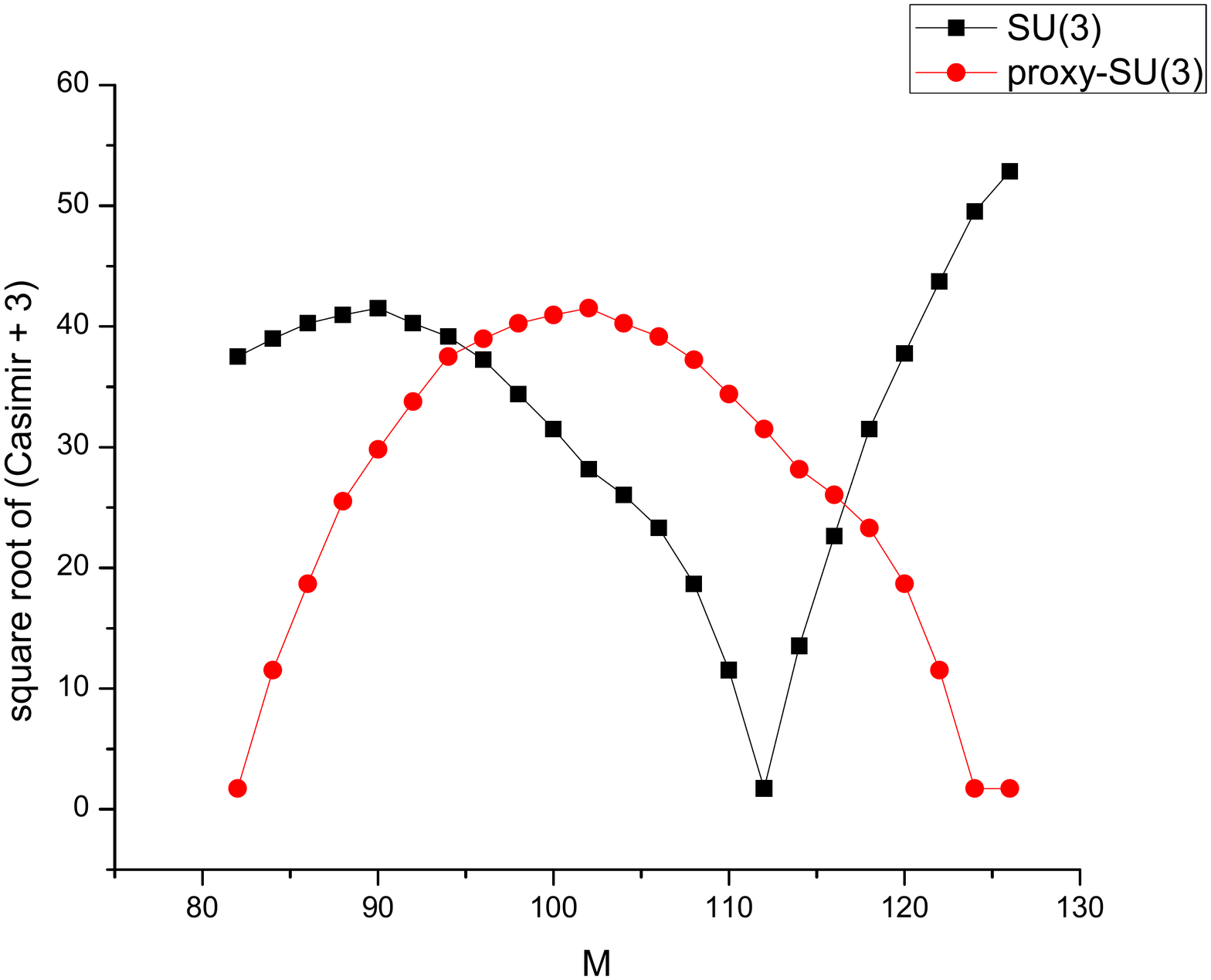}\hspace{1mm}
\includegraphics[width=55mm]{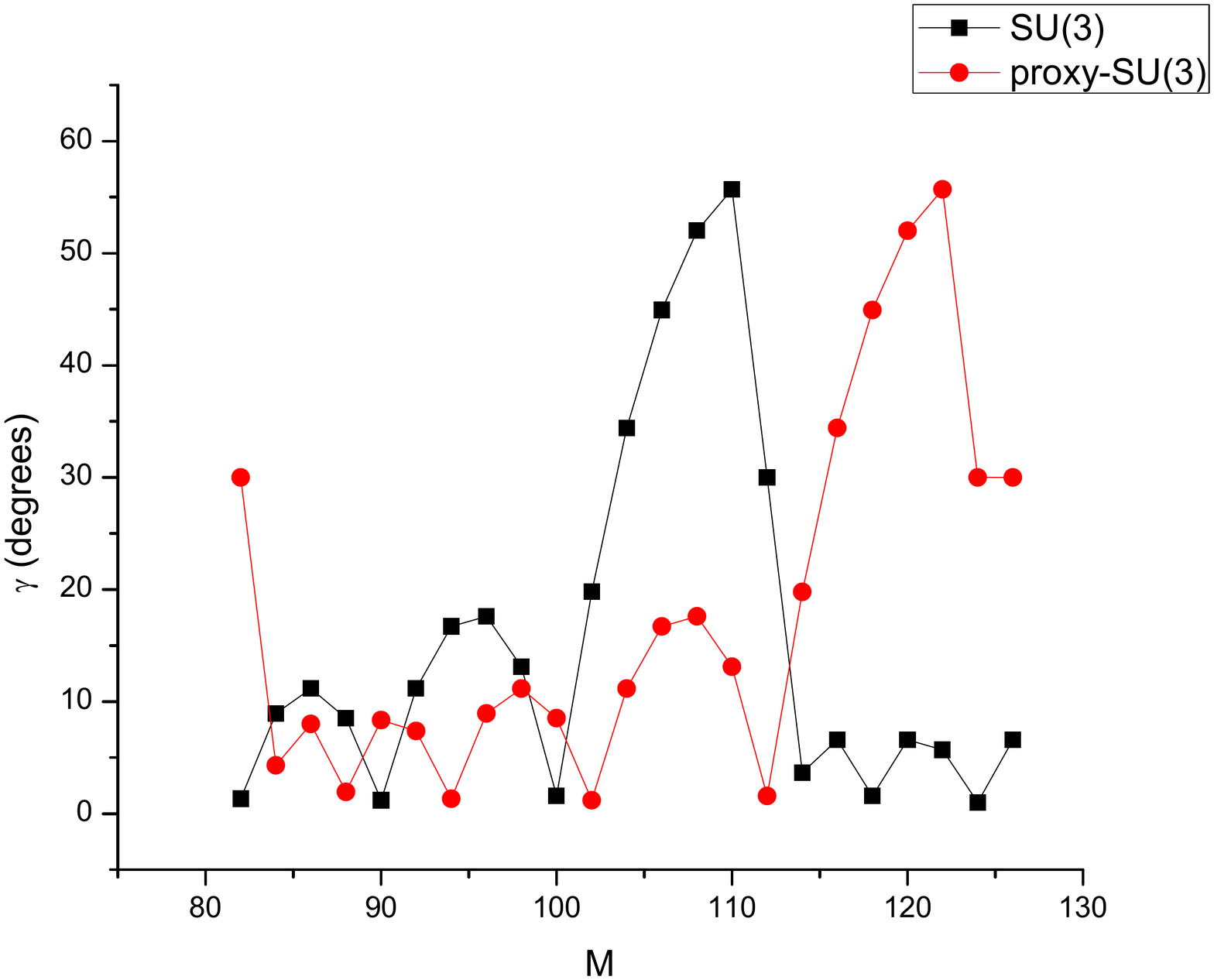}}
{\includegraphics[width=55mm]{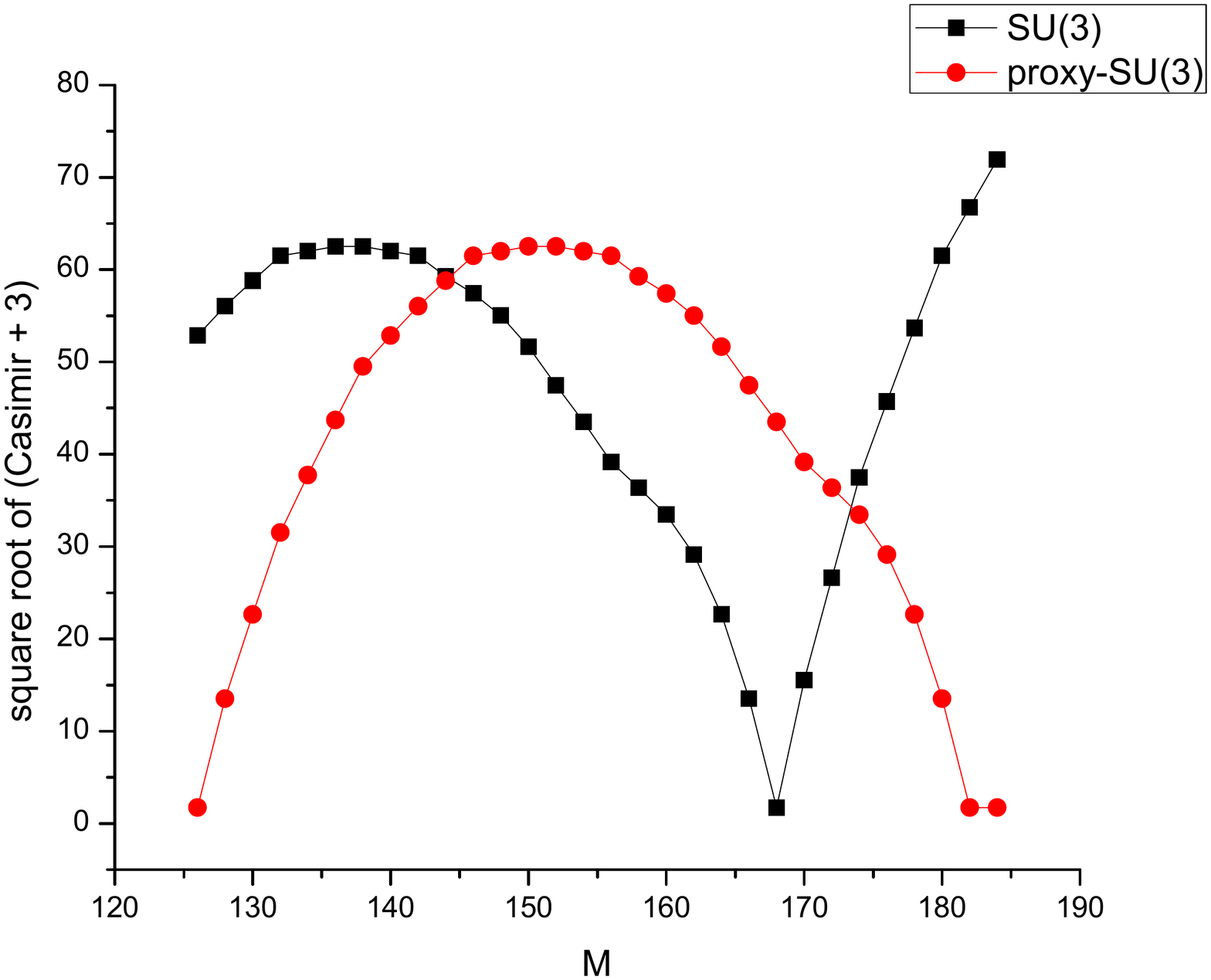}\hspace{1mm}
\includegraphics[width=55mm]{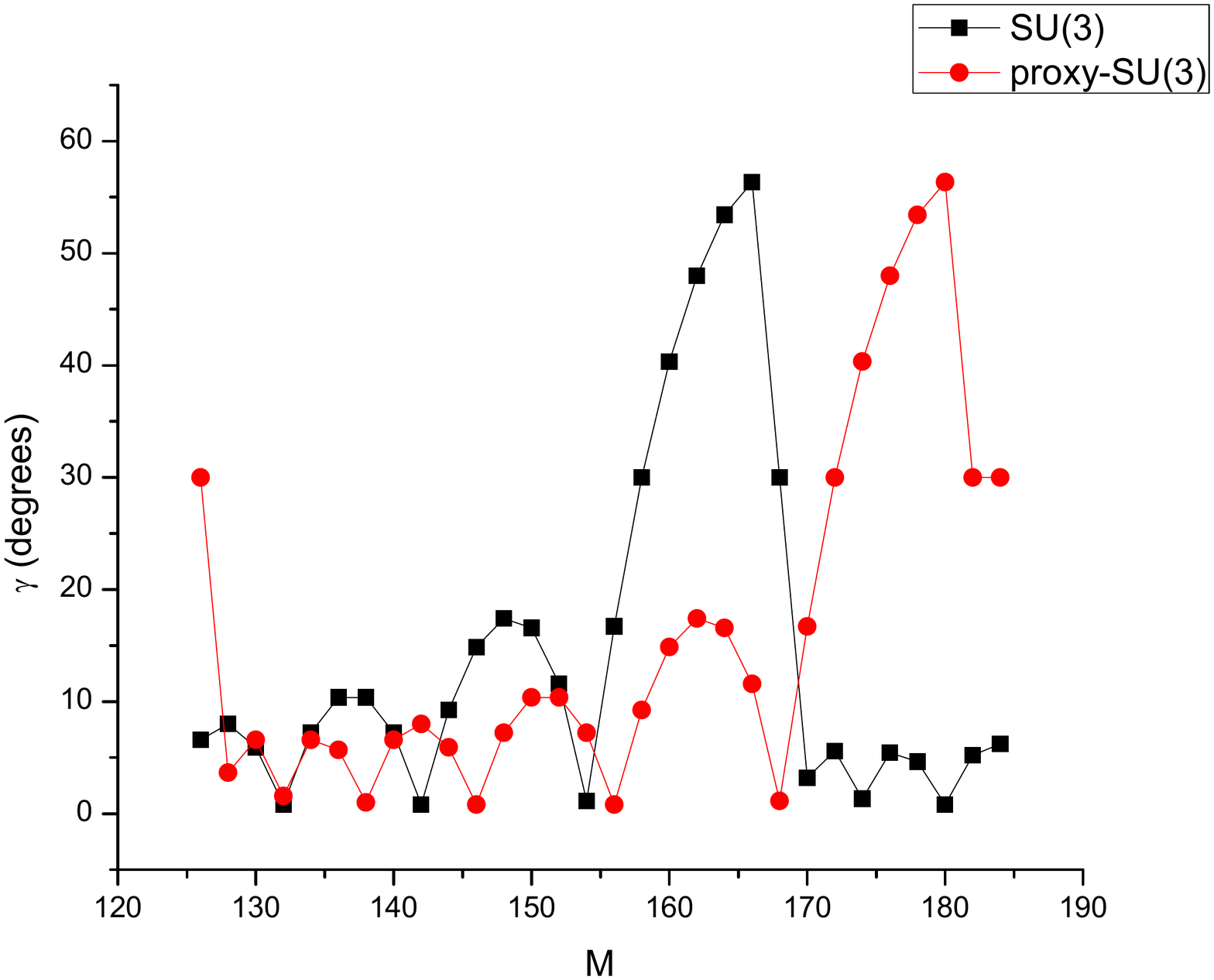}}

\caption{The quantity  $\sqrt{C_2(\lambda,\mu)+3}$, which is roughly proportional to the deformation parameter $\beta$ \cite{Castanos,Park,proxy2}, as well as the deformation parameter $\gamma$ 
(Eq. (\ref{g1})~) are plotted against the nucleon (proton or neutron) number $M$ for the 3D harmonic oscillator possessing the SU(3) symmetry, as well as for the proxy-SU(3) scheme. See Section \ref{coex} for further discussion.} 

\end{figure}

\end{document}